\documentclass[11pt, oneside]{article}   	% use "amsart" instead of "article" for AMSLaTeX format
\usepackage{geometry}                		% See geometry.pdf to learn the layout options. There are lots.
\usepackage{graphicx}				% Use pdf, png, jpg, or eps with pdflatex; use eps in DVI mode
\usepackage{amssymb}
\usepackage{amsmath}
\usepackage{subfig}
\usepackage{caption}
\usepackage{comment}

\usepackage[utf8]{inputenc}
\usepackage{amsmath}
\usepackage{graphicx}
\usepackage{float}	
\usepackage{amsmath}

\numberwithin{equation}{section}
\begin{document}
\title{Quantum Cosmology and Black Hole Interiors in Nonsupersymmetric String Theory and Canonical Gravity\\
}
\author{ Michael McGuigan \\
email: michael.d.mcguigan@gmail.com
}
\date{}
\maketitle
\begin{abstract}
In this paper we study black hole interior solutions and cosmologies in different dimensions using tools from canonical gravity and nonsupersymmetric string quantum cosmology. We find that the quantum wave functions associated with these solutions can be related to each other by a specific choice of variables. In a more realistic four dimensional setting  we combine canonical gravity and nonsupersymmetric string orbifold compactifications. We discuss the classical solutions and the corresponding wave functions involving inflation, the Higgs field, dark matter and hidden gauge sectors in these models. Finally we discuss string aspects of these models such as duality, massive modes, and nonperturbative approaches such as  Matrix theory and holography near the singularity.
\end{abstract}
\newpage

\section{Introduction}

Although much research has gone into classical and semiclassical analysis of black holes, especially near the horizon, it is also interesting to consider their behavior in the interior of the black hole near the singularity. In the interior, classical and semiclassical analysis can be supplemented by a fully quantum treatment. For the black holes treated in this paper, time and space exchange roles in the interior, and one has a picture of the interior as a time dependent cosmology. This is true both in classical and semiclassical descriptions and for a quantum description one has a  representation in terms of quantum cosmology. For quantum cosmology the Hamiltonian constraint of canonical gravity becomes an operator acting on states and its solution can represent the wave function of the black hole interior.

As one studies the black hole interior and approaches the singularity it is useful to embed the theory in a UV complete theory of gravity so that one can derive various coefficients of Effective Field Theories (EFTs) that are used for computations. This is similar to how more exact treatments of low energy phenomenology use EFTs to do computations with their parameters determined from a more fundamental description. For example this is how Fermi's four fermion effective interaction arises from the Weinberg-Salam model or how four quark operators can be derived from the standard model. In this paper we will take nonsupersymmetric string theory as the UV complete theory that gives rise to the low energy effective theories we use. There has been a rise of interest in nonsupersymmetric string theory as evidence of supersymmetry has not yet been seen at the LHC \cite{Alvarez-Gaume:1986ghj}
\cite{Dixon:1986iz}
\cite{Blaszczyk:2014qoa}
\cite{Abel:2015oxa}
\cite{Ashfaque:2015vta}
\cite{McGuigan:2019gdb}. Nonsupersymmetric string theory represents an alternative to supersymmetric string theory which preserves the  UV properties of string theory. It has compactifications with Higgs like fields and hidden sector which can play the role of dark matter as well as providing an explanation of why superpartners have not been seen in accelerators. Of course nonsupersymmetric string theory has at least the same difficulties as superstring theory, such as potential nonlocality, acausality and difficulty with time dependent backgrounds, an issue which is of great importance in cosmology. In addition it has problems related to stability and tadpoles. Thus it is a great challenge to construct a realistic nonsupersymmetric string theory with potential implications for inflationary and dark matter cosmology.

In this paper we explore several examples of black hole interiors and quantum cosmologies. In section two we study the black hole interior of a 2d Witten black hole associated with an exactly soluble nonsupersymmetric string CFT. In section three we study the simplest 4d black hole interior associated with the Schwarzschild solution and Kantowski-Sachs quantum cosmology. In section four we study a quantum cosmology associated to nonsupersymmetric string theory and show how it relates to the previous two solutions we considered. In section five we consider a more realistic setting for nonsupersymmetric string cosmology involving inflationary potentials and dark matter fields originating from orbifold compactifications of the nonsupersymmetric string. In all cases we discuss classical, semiclassical and fully quantum treatments of the black hole interiors and how different choices of variables of canonical gravity in minisuperspace relate to each other, and can be used to describe the quantum state of the cosmology or black hole interior near the singularity. Finally in section six we state the main conclusions and directions for future research.

%\section{Topology Change}

\section{2d Quantum Cosmology and Black hole Interior}

The simplest model of a black hole interior is related to the Witten black hole of 2d dilaton gravity \cite{Witten:1991yr}. This has the additional nice feature of having an exact conformal field theory associated with the solution related to a non-compact WZW model \cite{Dijkgraaf:1991ba}. The effective action for the 2d model involves the metric and dilaton and is given by:
\begin{equation}S = \int {{d^2}x\sqrt { - g} } {e^{ - 2\phi }}\left( {R + 4{{\left( {\nabla \phi } \right)}^2} + \frac{1}{{{\ell ^2}}}} \right)\end{equation}
where $-\frac{1}{\ell^2}$ is the negative cosmological constant that is necessary to have a black hole solution in 2d. As mentioned above the black hole interior can be described by a time dependent solution. This is because the black hole exterior can be described by static spatial dependent metric, and time and space are interchanged in the black hole interior. We introduce the metric ansatz used in minisuperspace canonical gravity given by:
\begin{equation}d{s^2} =  - {N^2}(t)d{t^2} + {a^2}(t)d{x^2}\end{equation}
where $N(t)$ is the lapse field and $a(t)$ represents the scale factor for the $x$ direction ( which was the time direction in the exterior). Using this ansatz and including a boundary term involving the extrinsic curvature to cancel second time derivatives, the Lagrangian becomes:
\begin{equation}L = \left( {4\frac{{\dot a\dot \phi }}{{aN}} - 4\frac{{{{\dot \phi }^2}}}{N} + \frac{N}{{{\ell ^2}}}} \right){e^{ - 2\phi }}\end{equation}
The solution to the Euler-Lagrange equations from this Lagrangian associated with the black hole interior is \cite{Tseytlin:1991xk}
\cite{Perry:1993ry}
\cite{Cadoni:1994av}:
\begin{equation}d{s^2} =  - d{t^2} + {\tan ^2}(t/2\ell )d{x^2}\end{equation}
so that:
\begin{align}
&N(t) = 1 \nonumber\\
&a(t) = \tan (t/2\ell)\nonumber \\
&e^{ - \phi(t)  } = \sqrt{2M} \cos (t/2\ell ) 
\end{align}
These solutions are plotted in figure 1 and 2. The horizon of the black hole corresponds to $t=0$ where $a$ vanishes and the singularity of the black hole corresponds to $t=\pi \ell$ where $a$ diverges. One can continue the solution beyond $t=\pi \ell$ but  what happens at or near the singularity cannot be addressed from the solutions to the action (2.1). One needs to embed the theory in a UV complete theory to address this, or use a canonical formulation of gravity to move beyond classical solutions and discuss wave functions, or ideally both. To pursue the canonical gravity approach one forms the Hamiltonian constraint by varying the Lagrangian with respect to the lapse variable $N$. For the 2d dilaton gravity this becomes:
\begin{equation}H = \left( { - 4\frac{{\dot a\dot \phi }}{{a{N^2}}} + 4\frac{{{{\dot \phi }^2}}}{{{N^2}}} + \frac{1}{{{\ell ^2}}}} \right){e^{ - 2\phi }} = 0\end{equation}
then obtaining the canonical momentum associated with $a$ and $\phi$ from the Lagrangian we have:
\[{p_a} = 4\frac{{\dot \phi }}{{aN}}\]
\begin{equation}{p_\phi } =  - 8\frac{{\dot \phi }}{N}\end{equation}
Now expressing the Hamiltonian constraint in terms of canonical momentum :
\begin{equation}H =  - \frac{{{e^{2\phi }}}}{{4a}}{p_a}{p_\phi } - \frac{{{e^{2\phi }}}}{4}p_a^2 + \frac{1}{{{\ell ^2}}} = 0\end{equation}
%\[\delta {s^2} =  - 4{a^2}{e^{ - 2\phi }}d{\phi ^2} + 4a{e^{ - 2\phi }}dad\phi \]
%\[L = \left( {4\frac{{\dot a\dot \phi }}{{aN}} - 4\frac{{{{\dot \phi }^2}}}{N} + \frac{N}{{{\ell ^2}}}} \right){e^{ - 2\phi }}\]
The (Wheeler-DeWitt (WDW) equation for the wave function is then given by:
\begin{equation}H \psi = ( - \frac{{{e^{2\phi }}}}{{4a}}{p_a}{p_\phi } - \frac{{{e^{2\phi }}}}{4}p_a^2 + \frac{1}{{{\ell ^2}}}) \psi = 0\end{equation}
To represent this equation as a second order differential equation one uses the substitution:
\[{p_a} =  - i\frac{\partial }{{\partial a}}\]
\begin{equation}{p_\phi} =  - i\frac{\partial }{{\partial \phi}}\end{equation}
There are three types of solutions to this WDW equation we will consider. The first is a WKB type solution best expressed in terms of the original $(a,\phi)$ coordinates, the second is a plane wave type solution based on lightcone coordinates in minisuperspace, and the third is a Rindler type wave function based on Rindler coordinates in minisuperspace. We will consider each of these in turn. 

\subsection*{WKB wave functions using $(a,\phi)$ minisuperspace}

For the WKB like wave function one first forms a Mass operator from the canonical momentum and minisuperspace coordinates given by:
\begin{equation}M = \frac{{{e^{2\phi }}\ell p_a^2}}{8} + \frac{{{e^{ - 2\phi }}}}{{2\ell }}\end{equation}
One then solves this equation for $p_a$ to obtain
\begin{equation}{p_a} = \sqrt {\frac{{8M}}{{{e^{2\phi }}\ell }} - \frac{{8{e^{ - 4\phi }}}}{{2{\ell ^2}}}} \end{equation}
The WKB like wave function is then of the form 
\begin{equation}{\psi _{WKB}} = \frac{1}{{\left| {{p_a}} \right|}}\exp \left( { \pm ia{p_a}} \right)\end{equation}
Defining $W_M(\phi)$ by:
\begin{equation}{W_M}(\phi ) = \sqrt {\frac{{8M}}{{{e^{2\phi }}\ell }} - \frac{{8{e^{ - 4\phi }}}}{{2{\ell ^2}}}} \end{equation}
one writes the WKB like wave function as:
\begin{equation}{\psi _M}(a,\phi ) = \frac{1}{{{W_M}(\phi )}}\exp \left( { \pm ia{W_M}(\phi )} \right)\end{equation}
These solutions have the structure:
\begin{equation}{\psi _M} = f_M\exp \left( { \pm i{S_M}} \right)\end{equation}
with $S_M$ a solution to the Hamilton-Jacobi equation and $f_M$ a WKB prefactor. These solutions can be used in a semiclassical approximation and a superposition of the form:
\begin{equation}\psi  = \sum\limits_M {({A_M}f_M\exp \left( { + i{S_M}} \right)}  + {B_M}f_M\exp \left( { - i{S_M}} \right))\end{equation}
This superposition of WKB like wave functions can be used to represent wavepacket states which track the classical solutions.

\subsection*{Plane wave like wave functions using lightcone minisuperspace coordinates}

For the plane wave like solutions one defines lightcone coordinates in minisuperspace by
\begin{align}
&u = {a^2}{e^{ - 2\phi }}\nonumber\\
&v = {e^{ - 2\phi }}
\end{align}
In these coordinates the Hamiltonian constraint is very simple and is given by:
\begin{equation}{H=p_u}{p_v} - \ell^{-2} = 0\end{equation}
The solution to the WDW equation $H\psi=0$ is then:
\begin{equation}{\psi _{LC}} = \exp \left( { \pm i({p_u}u + {p_v}v)} \right)\end{equation}
Defining $K$ by $p_u=K$ and $p_v = \frac{1}{K \ell^2}$ and expressing the wave function in the $(a,\phi)$ variables we have:
\begin{equation}\psi _K^{ \pm (2)}(a,\phi ) = \exp \left( { \pm \frac{i}{2}\left( { - K{a^{2}}{e^{-2\phi }} - \frac{{{e^{-2\phi }}}}{K \ell^2}} \right)} \right)\end{equation}
The general solution can be written as a superposition of the form:
\begin{equation}\psi (a,\phi ) = \sum\limits_K {\left( {A_K^{(2)}\psi _K^{ + (2)}(a,\phi ) + B_K^{(2)}\psi _K^{ - (2)}(a,\phi )} \right)} \end{equation}

\subsection*{Rindler like wave solutions using Rindler minisuperspace coordinates}

Using the lightcone coordinates one can define Rindler like coordinates  $(\xi,\eta)$ in minisuperspace by:
\begin{align}
&u = {e^\xi }{e^\eta }\nonumber \\
&v = {e^\xi }{e^{-\eta }}
\end{align}
For the case of 2d nonsupersymmetric string theory this gives:
\begin{align}
&{e^\eta } = a\nonumber \\
&{e^\xi } = {a}{e^{-2\phi }}
\end{align}
The Hamiltonian constraints in these coordinates is given by:
\begin{equation}H = {e^{ - 2\xi }}\left( { - p_\eta ^2 + p_\xi ^2} \right) + \ell^{-2} = 0\end{equation}
The solutions to the WDW equation in terms of the Rindler like $(\xi,\eta)$ minisuperspace variables are the Rindler wave functions which can be expressed in terms of the modified Bessel function of the second kind $K_\nu(z)$ through:
\begin{equation}{\psi _k}^{ \pm (3)}(\eta ,\xi ) = {K_{ik}}\left( {\ell^{-1}  {e^\xi }} \right){e^{ \pm ik\eta }}\end{equation}
in terms of the original $(a,\phi)$ minisuperspace coordinates this is:
\begin{equation}\psi _k^{ \pm (3)}(a,\phi ) = {\left( a \right)^{ \pm ik}}{K_{ik}}\left( \ell^{-1}{a{e^{ - 2\phi }}} \right)\end{equation}
The general solution can be written as a superposition of the form:
\begin{equation}\psi (a,\phi ) = \sum\limits_k {\left( {A_k^{(3)}\psi _k^{ + (3)}(a,\phi ) + B_k^{(3)}\psi _k^{ - (3)}(a,\phi )} \right)} \end{equation}

\begin{figure}
\centering
  \includegraphics[width = .75 \linewidth]{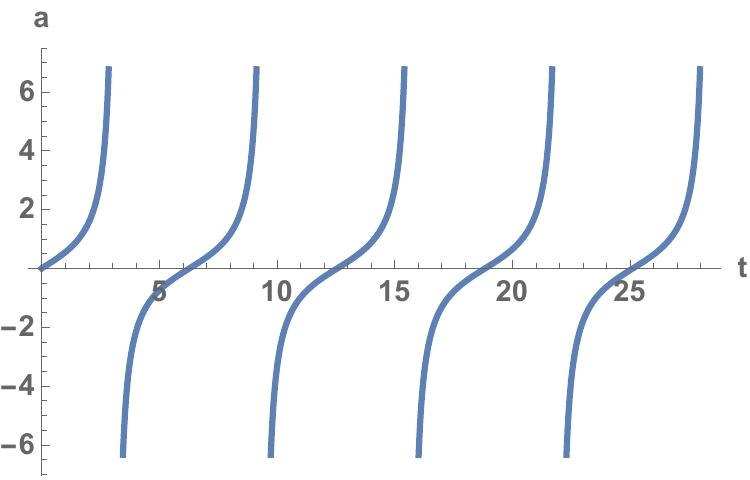}
  \caption{Scale factor $a$ for the 2d black hole interior solution drawn for $\ell=1$. The scale factor vanishes at the horizon for $t=0$ and diverges at the black hole singularity for $t=\pi \ell$.}
  \label{fig:Radion Potential}
\end{figure}

\begin{figure}
\centering
  \includegraphics[width = .75 \linewidth]{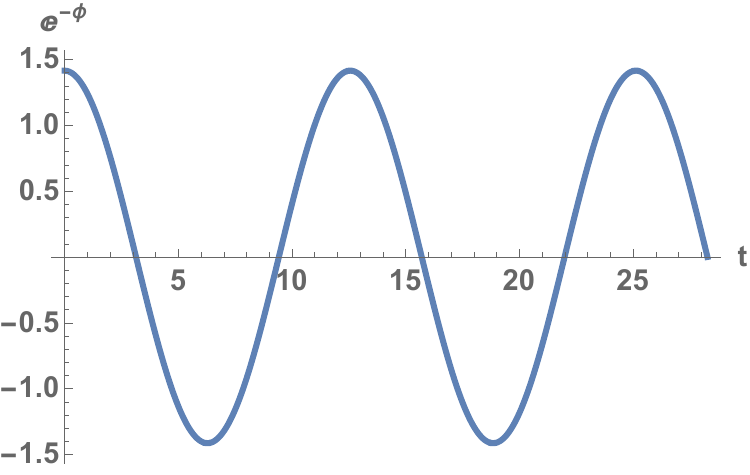}
  \caption{Exponential of minus the dilaton field for the 2d black hole interior solution drawn for $\ell=1$ and $M=1$.The dilaton field vanishes at the horizon for $t=0$ and diverges at the black hole singularity for $t=\pi \ell$.}
  \label{fig:Radion Potential}
\end{figure}

%\section{3d gravity and two sided black holes}

\section{4d Quantum Cosmology and Black hole Interior}

For 4d gravity the black hole interior can be treated in a similar manner to the 2d dilaton gravity considered above. One starts with the Einstein-Hilbert action given by
\begin{equation}S = \frac{1}{2}\int {{d^4}x\sqrt { - g} }  R \end{equation}
The black hole interior solutions are time dependent. This is because like in the 2d case the exterior solution can be described by spatial dependent static metric and time and space are interchanged in the black hole interior. Using the metric ansatz for 4d minisuperspace canonical gravity:
\begin{equation}d{s^2} =  - {N^2}(t)d{t^2} + {a^2}(t)d{x^2} + {b^2}(t)d\Omega _2^2\end{equation}
with $N(t)$ the lapse field and $a(t)$ are scale factors for the $x$ direction (which was the time direction in the exterior) and $b(t)$ is the scale factor associated with a two sphere with $d\Omega _2^2$ the metric on the unit two sphere. Using this ansatz and including a boundary term involving the extrinsic curvature which removes second time derivatives the Lagrangian becomes:
\begin{equation}L = \frac{1}{2}\left( { - \frac{{2b\dot a\dot b}}{N} - \frac{{a{{\dot b}^2}}}{N} + aN} \right)\end{equation}
The black hole interior solution to the Euler-Lagrange equations from this Lagrangian is given by \cite{Peeters:1994jz}:
\begin{equation}d{s^2} =  - 4{M^2}{\cos ^4}(t/2)d{t^2} + {\tan ^2}(t/2)d{x^2} + 4{M^2}{\cos ^4}(t/2)d\Omega _2^2\end{equation}
 The $N,a,b$ solutions are given by:
\begin{align}
 &N(t) = 2M{\cos ^2}(t/2)\nonumber \\
&a(t) = \tan (t/2)\nonumber \\
&b(t) = 2M{\cos ^2}(t/2)   
\end{align}
These solution are plotted in figures 3,4,5,6. The horizon corresponds to $t=0$ and singularity to $t=\pi$ where $a$ diverges. One can continue the solution beyond $t=\pi$ but the interesting region near the singularity requires a UV complete theory or a quantization beyond classical solutions to address the resolution of the singularity. To pursue a canonical gravity approach one forms the Hamiltonian constraint as above by varying the Lagrangian with respect to the Lapse variable $N$. For 4d gravity one obtains
\begin{equation}H = \frac{{2b\dot a\dot b}}{{{N^2}}} + \frac{{a{{\dot b}^2}}}{{{N^2}}} + a\end{equation}
The canonical momentum associated with $a$ and $b$ from the Lagrangian are given by:
\begin{align}
&{p_a} =  - \frac{{b\dot b}}{N}\nonumber \\
&{p_b} =  - \frac{{b\dot a + a\dot b}}{N}
\end{align}
which for the interior solution are given by:
\begin{align}
&{p_a} = 2M\cos (t/2)\sin (t/2)\nonumber \\
&{p_b} =  - \frac{1}{2}\left( {1 - {{\tan }^2}(t/2)} \right)
\end{align}
Now expressing the Hamiltonian constraint in terms of canonical momentum we have:
\begin{equation}H= - \frac{{p_a^2}}{{{b^2}}} + \frac{{2{p_a}{p_b}}}{{ab}} + 1 = 0\end{equation}
The WDW for the equation for the wave function is given by:
\begin{equation}H\psi = \left(- \frac{{p_a^2}}{{{b^2}}} + \frac{{2{p_a}{p_b}}}{{ab}} + 1\right)\psi = 0\end{equation}
This equation is similar to the WDW equation for Kantowski-Sachs cosmology of $S^1 \times S^2$ spatial topology \cite{Conradi:1994yy}
\cite{Uglum:1992nc}
\cite{Louko:1988ia}
\cite{Fanaras:2022twv}
\cite{Fanaras:2021awm}
\cite{Gates:1998py}
\cite{Kadoyoshi:1997wt}
\cite{Nojiri:1999iv}
\cite{Bronnikov:2008by}
\cite{Halliwell:1990tu}
\cite{Unruh:1998nn}
\cite{Chakraborty:1991mf}. To represent this equation as a second order differential equation one makes the substitution:
\[{p_a} =  - i\frac{\partial }{{\partial a}}\]
\begin{equation}{p_b} =  - i\frac{\partial }{{\partial b}}\end{equation}
As in the 2d gravity case  there are three types of solutions to this WDW equation we will consider. The first is a WKB type solution best expressed in terms of the orginal $(a,b)$ coordinates, the second is a plane wave type solution based on lightcone coordinates in minisuperspace and the third is a Rindler type wave function based on Rindler coordinates in minisuperspace. We will again consider each of these in turn. 
%\[{G_{IJ}} = \left( {\begin{array}{*{20}{c}}
%0&{4\frac{b}{a}}\\
%{4\frac{b}{a}}&4
%\end{array}} \right)\]

%Another form of the black hole interior solution in the gauge $Na=1$ is given by:
%\begin{align}
%&N(t) = {\left( {\frac{{2M}}{t} - 1} \right)^{1/2}}\nonumber \\
%&a(t) = {\left( {\frac{{2M}}{t} - 1} \right)^{1/2}}\nonumber \\
%&b(t) = t
%\end{align}
%which we plot in figure 2. The big crunch in this form of the interior solution occurs at $t=0$. 
%\[{p_a} = 2M\cos (t/2)\sin (t/2)\]
%\[{p_b} =  - \frac{1}{2}\left( {1 - {{\tan %}^2}(t/2)} \right)\]

\subsection*{WKB wave function using $(a,b)$ minisuperspace}

For the WKB like wave function one first forms a mass operator from the canonical momentum and minisuperspace coordinates as:
\begin{equation}M = \frac{{p_a^2}}{{2b}} + \frac{b}{2}\end{equation}
One then solves this equation for $p_a$ to obtain:
\begin{equation}p_a = \sqrt {2Mb - {b^2}} \end{equation}
The WKB like wave function is then of the form 
\begin{equation}{\psi _{WKB}} = \frac{1}{{\left| {{p_a}} \right|}}\exp \left( { \pm ia{p_a}} \right)\end{equation}
Defining $W_M(b)$ by:
\begin{equation}{W_M}(b) = \sqrt {2Mb - {b^2}} \end{equation}
one writes the WKB like wave function as:
\begin{equation}{\psi _M}(a,\phi ) = \frac{1}{{{W_M}(\phi )}}\exp \left( { \pm ia{W_M}(\phi )} \right)\end{equation}
in agreement with the structure of the wave functions found in \cite{Conradi:1994yy}. These solutions have the structure:
\begin{equation}{\psi _M} = f_M\exp \left( { \pm i{S_M}} \right)\end{equation}
with $S_M$ a solution to the Hamilton-Jacobi equation and $f_M$ a WKB prefactor. The solutions can be used in a semiclassical approximation. General superpositions are expressed as:
\begin{equation}\psi  = \sum\limits_M {({A_M}f_M\exp \left( { + i{S_M}} \right)}  + {B_M}f_M\exp \left( { - i{S_M}} \right))\end{equation}
These superpositions of WKB like wave functions can be used to represent wavepacket states which track the classical solutions.

\subsection*{Plane wave like wave functions using lightcone minisuperspace coordinates}

For the plane wave like solutions one defines lightcone coordinates in minisuperspace by
\begin{align}
&u = a^2 b\nonumber \\
&v = b
\end{align}
The black hole interior solution in the Lightcone $(u,v)$ minsuperspace coordinates is shown in figure 7.
In these coordinates the Hamiltonian constraint is very simple and is given by:
\begin{equation}{H=p_u}{p_v} - 1 = 0\end{equation}
The solution to the WDW equation $H\psi=0$ is then:
\begin{equation}{\psi}  = \exp \left( { \pm i({p_u}u + {p_v}v)} \right)\end{equation}
Defining $K$ by $p_u=K$ and $p_v = \frac{1}{K}$ and expressing the wave function in the $(a,b)$ variables we have:
\begin{equation}\psi _K^{ \pm (2)}(a,b ) = \exp \left( { \pm \frac{i}{2}\left( { - K{a^2 b} - \frac{{b}}{K}} \right)} \right)\end{equation}
The general solution can be written as a superposition of the form:
\begin{equation}\psi (a,b ) = \sum\limits_K {\left( {A_K^{(2)}\psi _K^{ + (2)}(a,b ) + B_K^{(2)}\psi _K^{ - (2)}(a,b)} \right)} \end{equation}

\subsection*{Rindler like wave solutions using Rindler minisuperspace coordinates}

As in the 2d case using the lightcone coordinates one can define Rindler like coordinates  $(\xi,\eta)$ in minisuperspace by:
\begin{align}
&u = {e^\xi }{e^\eta }\nonumber \\
&v = {e^\xi }{e^{-\eta }}
\end{align}
For the case of 4d gravity this gives:
\[{e^\eta } = a\]
\begin{equation}{e^\xi } = ab\end{equation}
The black hole interior solution in the Rindler $(\xi,\eta)$ minisuperspace coordinates is shown in figure 8.
The Hamiltonian constraints in these coordinates is given by:
\begin{equation}H = {e^{ - 2\xi }}\left( { - p_\eta ^2 + p_\xi ^2} \right) + 1 = 0\end{equation}
The solutions to the WDW equation $H\psi=0$ in terms of the Rindler like $(\xi,\eta)$ minisuperspace variables are the Rindler wave functions. These can be expressed in terms of the modified Bessel function of the second kind $K_\nu(z)$ through:
\begin{equation}{\psi _k}^{ \pm (3)}(\eta ,\xi ) = {K_{ik}}\left( { {e^\xi }} \right){e^{ \pm ik\eta }}\end{equation}
In terms of the original $(a,b)$ minisuperspace coordinates this is:
\begin{equation}\psi _k^{ \pm (3)}(a,b ) = {\left( {{a}} \right)^{ \pm ik}}{K_{ik}}\left( ab\right)\end{equation}
The general solution can be written as a superposition of the form:
\begin{equation}\psi (a,\phi ) = \sum\limits_k {\left( {A_k^{(3)}\psi _k^{ + (3)}(a,\phi ) + B_k^{(3)}\psi _k^{ - (3)}(a,\phi )} \right)} \end{equation}

\begin{figure}
\centering
  \includegraphics[width = .75 \linewidth]{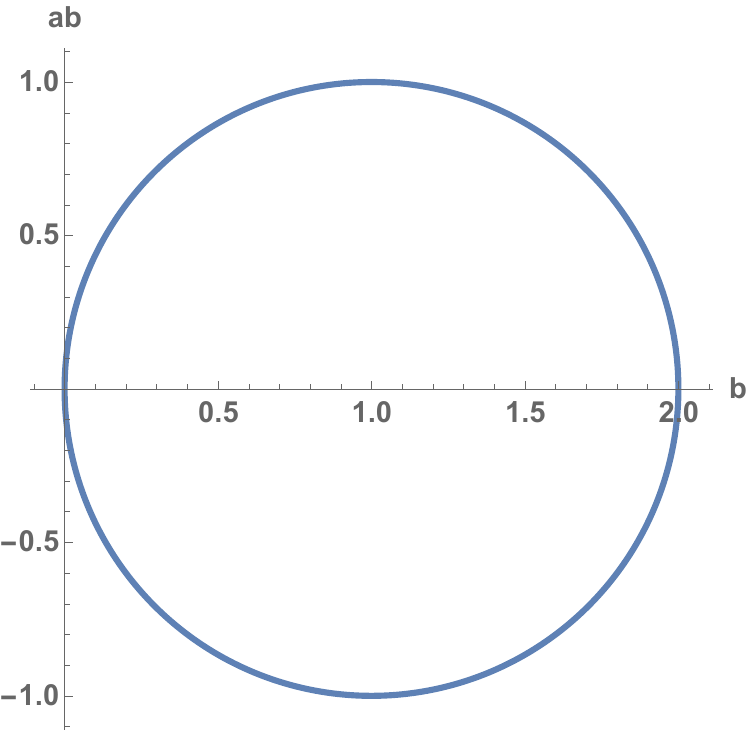}
  \caption{Product of scale factors $ab$ versus $b$ for the 4d black hole interior solution with $M=1$.The product of scale factors  $ab$ goes to zero and $b$ goes to $2M$ at the horizon. $ab$ goes to zero and $b$ goes to zero at the black hole singularity.}
  \label{fig:Radion Potential}
\end{figure}

\begin{figure}
\centering
  \includegraphics[width = .75 \linewidth]{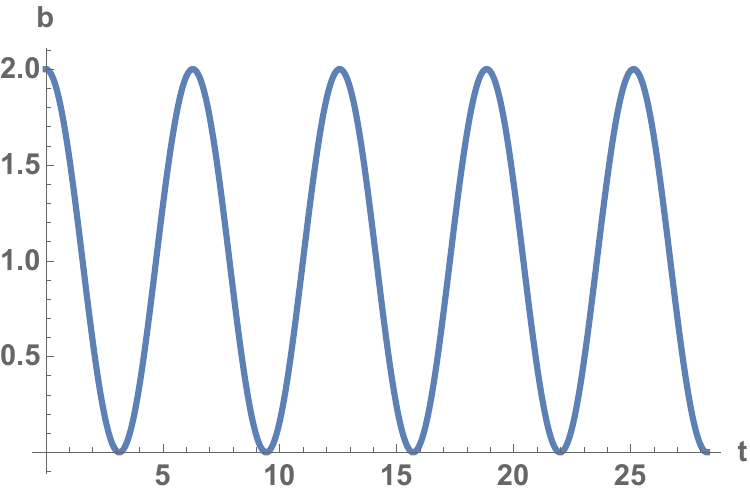}
  \caption{Scale factor $b$ for the 4d black hole interior solution with $M=1$.The scale factor $b$ goes to $2M$ at the black hole horizon for $t=0$ and goes to zero at the black hole singularity for $t=\pi$.}
\end{figure}

\begin{figure}
\centering
  \includegraphics[width = .75 \linewidth]{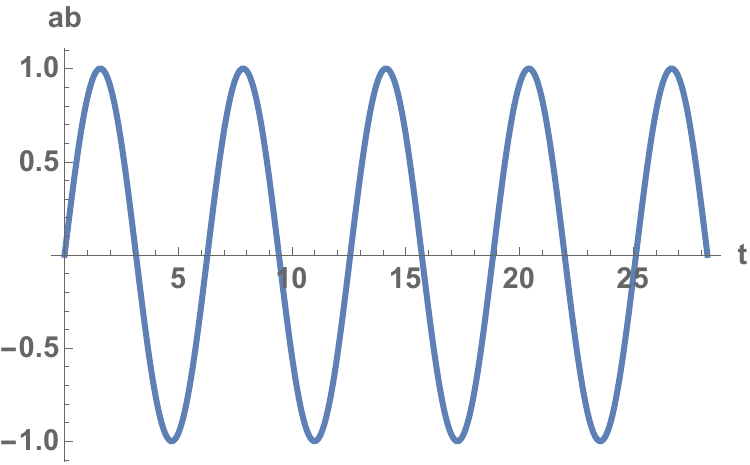}
  \caption{Product of scale factors $ab$ for the 4d black hole interior solution with $M=1$.The product of scale factors $ab$ goes to zero at the black hole horizon for $t=0$ and goes to zero at the black hole singularity for $t=\pi$.}
  \label{fig:Radion Potential}
\end{figure}

\begin{figure}
\centering
  \includegraphics[width = .75 \linewidth]{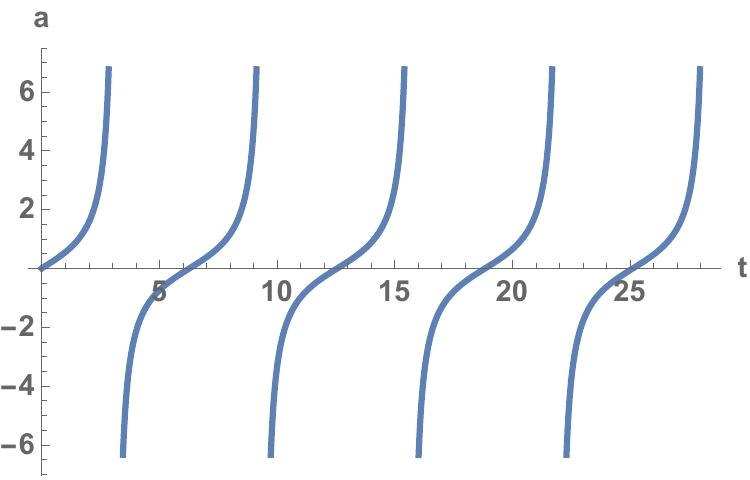}
  \caption{Scale factor $a$ for the 4d black hole interior solution with $M=1$.The scale factor $a$ goes to zero at the black hole horizon for $t=0$ and goes to infinity at the black hole singularity for $t=\pi$.}
  \label{fig:Radion Potential}
\end{figure}

\begin{figure}
\centering
  \includegraphics[width = .5 \linewidth]{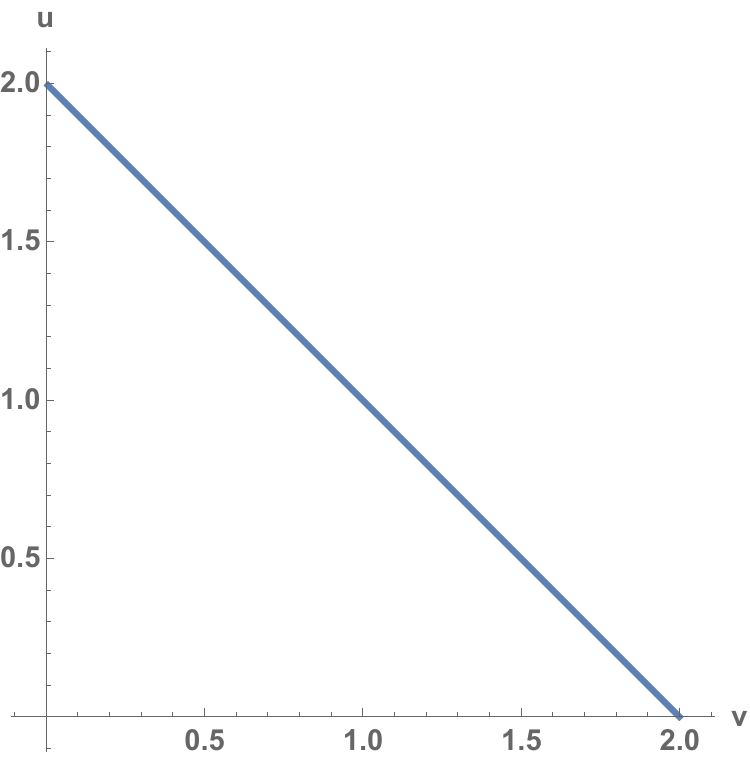}
  \caption{$u$ versus $v$ for the 4d black hole interior solution which follows a straight geodesic in lightcone minisuperspace coordinates. }
  \label{fig:Radion Potential}
\end{figure}

\begin{figure}
\centering
  \includegraphics[width = .25 \linewidth]{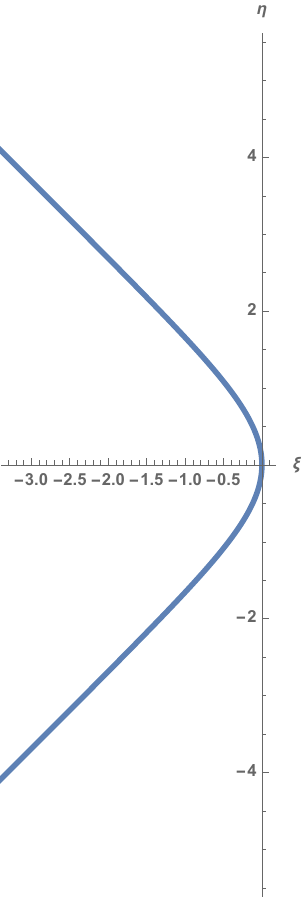}
  \caption{$\eta$ versus $\xi$ for the 4d black hole interior solution which follows a curved geodesic in Rindler minisuperspace coordinates.}
  \label{fig:Radion Potential}
\end{figure}

\section{10d nonsupersymmetric string quantum cosmology}

Nonsupersymmetric string theory is of great interest as the LHC has yet to discover supersymmetry and as it represents a UV complete theory that includes gravity without superpartners. The theory can be compactified on orbifolds to obtain realistic models containing Higgs bosons, dark matter candidates and hidden sectors which have potential observational signatures in particle collider and astrophysical measurements. For cosmology it is simpler to study the 10d theory at first and we will consider the compactified nonsupersymmetric string theory in the next section. The 10d nonsupersymmetric string effective action including the metric and dilaton in the Einstein frame is given by: 
\begin{equation}S = \int d^{10}x\frac{1}{2}\sqrt { - g} \left( {R - \frac{1}{2}{{\left( {\nabla \phi } \right)}^2} - 2\lambda {e^{5\phi /2}}} \right)\end{equation}
For a cosmological solution we choose an ansatz where the scale factor and dilaton depend on time given by:
\[d{s^2} =  - {N^2}(t)d{t^2} + {a^2}(t)\sum\limits_{i = 1}^9 {d{x_i}^2} \]
\begin{equation}{e^\phi } = {e^{\phi (t)}}\end{equation}
Using this ansatz and including a boundary term involving the extrinsic curvature which removes second time derivatives the Lagrangian becomes:
\begin{equation}L =  - 36\frac{{{a^7}{{\dot a}^2}}}{N} + \frac{1}{4}\frac{{{a^9}{{\dot \phi }^2}}}{N} - \lambda {a^9}{e^{5\phi /2}}N\end{equation}
The cosmological solution to the Euler-Lagrange equations from this Lagrangian is given by \cite{Dudas:2000ff}:
\begin{align}
&N(t) = {\left( {\sin (\sqrt \lambda  t)} \right)^{ - 5/8}}{\left( {\cos (\sqrt \lambda  t)} \right)^{-5/2}}\nonumber\\
&a(t) = {\left( {\sin (\sqrt \lambda  t)} \right)^{1/24}}{\left( {\cos (\sqrt \lambda  t)} \right)^{ - 1/6}}\nonumber\\
&{e^{{\phi (t)}-\phi_0}} = {\left( {\sin (\sqrt \lambda  t)} \right)^{1/2}}{\left( {\cos (\sqrt \lambda  t)} \right)^2}
\end{align}
These solutions are plotted in figures 9 and 10. The curvature singularity in the metric corresponds to time $t=0$ and  $t=\pi/(2\sqrt{\lambda})$ where $a$ diverges.  To pursue a canonical gravity approach one forms the Hamiltonian constraint as above by varying the Lagrangian with respect to the lapse variable $N$. For 4d gravity one obtains
\begin{equation}H =  - 36\frac{{{a^7}{{\dot a}^2}}}{{{N^2}}} + \frac{1}{4}\frac{{{a^9}{{\dot \phi }^2}}}{{{N^2}}} - \lambda {a^9}{e^{5\phi /2}}\end{equation}
The canonical momentum associated with $a$ and $\phi$ from the Lagrangian are given by:
\begin{align}
&{p_a} =  - \frac{{72{a^7}\dot a}}{N}\nonumber \\
&{p_\phi } = \frac{{{a^9}\dot \phi }}{{2N}}
\end{align}
Now expressing the Hamiltonian constraint in terms of canonical momentum we have:
\begin{align}H =  - \frac{{p_a^2}}{{144{a^7}}} + \frac{{p_\phi ^2}}{{{a^9}}} + \lambda {a^9}{e^{5\phi /2}}=0\end{align}
The WDW for the equation for the wave function is given by:
\begin{align}H \psi = \left( - \frac{{p_a^2}}{{144{a^7}}} + \frac{{p_\phi ^2}}{{{a^9}}} + \lambda {a^9}{e^{5\phi /2}}\right)\psi=0\end{align}
To represent this equation as a second order differential equation one makes the substitution:
\[{p_a} =  - i\frac{\partial }{{\partial a}}\]
\begin{align}{p_\phi} =  - i\frac{\partial }{{\partial \phi}}\end{align}
Although this 10d nonsupersymmetric cosmological solution does not represent a black hole interior, its wave functions  can be represented using different minisuperspace coordinates as in the previous two cases. As in the 2d and 4d gravity case  there are different types of solutions to this WDW equation, We will consider a plane wave type solution based on lightcone coordinates in minsuperspace and the  a Rindler type wave function based on Rindler coordinates in minisuperspace. 
%\[H =  - \frac{{p_a^2}}{{144{a^7}}} + %\frac{{p_\phi ^2}}{{{a^9}}} + \lambda %{a^9{e^{5\phi /2}}\]
%\[{G_{IJ}} = \left( {\begin{array}{*{20}{c}}
%{ - 144{a^{16}}{e^{5\phi /2}}}&0\\
%0&{{a^{18}}{e^{5\phi /2}}}
%\end{array}} \right)\]
%\[{\left( {{\mathcal{G}^{ - 1}}} \right)^{IJ}} = \left( {\begin{array}{*{20}{c}}
%{ - \frac{1}{{144{a^{16}}{e^{5\phi /2}}}}}&0\\
%0&{\frac{1}{{{a^{18}}{e^{5\phi /2}}}}}
%\end{array}} \right)\]
%\[{\mathcal{P}_I} = \frac{1}{2}{\mathcal{G}_{IJ}}%{{\mathcal{\dot X}}^J}\]
%\[{\left( {{\mathcal{G}^{ - 1}}} \right)^{IJ}}%{\mathcal{P}_I}{\mathcal{P}_J} + \lambda  = 0\]
%Choosing the gauge 
%$N{e^{5\phi /4}} = 1$
%we have the exact cosmological solution of Dudas and Mourad\cite{Dudas:2000ff} written in the Einstein-frame given by:
%\begin{align}
%&N(t) = {\left( {\sin (\sqrt \lambda  t)} \right)^{ - 5/8}}{\left( {\cos (\sqrt \lambda  t)} \right)^{-5/2}}\nonumber\\
%&a(t) = {\left( {\sin (\sqrt \lambda  t)} \right)^{1/24}}{\left( {\cos (\sqrt \lambda  t)} \right)^{ - 1/6}}\nonumber\\
%&{e^{{\phi (t)}-\phi_0}} = {\left( {\sin (\sqrt \lambda  t)} \right)^{1/2}}{\left( {\cos (\sqrt \lambda  t)} \right)^2}
%\end{align}

\subsection*{Plane wave like wave functions using lightcone minisuperspace coordinates}

For the plane wave like solutions one defines lightcone coordinates in minisuperspace for 10d nonsupersymmetric string theory as:
\begin{align}
&u = {a^{24}}{e^{2\phi }}\nonumber \\
&v = {a^{ - 6}}{e^{\phi /2}}
\end{align}
In these coordinates the Hamiltonian constraint is very simple and is given by:
\begin{equation}{H=p_u}{p_v} - \lambda = 0\end{equation}
The solution to the WDW equation $H\psi=0$ is then:
\begin{equation}\psi  = \exp \left( { \pm i({p_u}u + {p_v}v)} \right)\end{equation}
Defining $K$ by $p_u=K$ and $p_v = \frac{\lambda}{K}$ and expressing the wave function in the $(a,b)$ variables
the WDW solution then takes the simple form:
\begin{equation}\psi _K^{ \pm (2)}(a,\phi ) = \exp \left( { \pm \frac{i}{2}\left( { - K{a^{24}}{e^{2\phi }} - \lambda \frac{{{a^{ - 6}}{e^{\phi /2}}}}{K}} \right)} \right)\end{equation}
The general solution can be written as a superposition of the form:
\begin{equation}\psi (a,\phi ) = \sum\limits_K {\left( {A_K^{(2)}\psi _K^{ + (2)}(a,\phi ) + B_K^{(2)}\psi _K^{ - (2)}(a,\phi )} \right)} \end{equation}

\subsection*{Rindler like wave solutions using Rindler minisuperspace coordinates}

As in the 2d and 4d case using the lightcone coordinates one can define Rindler like coordinates  $(\xi,\eta)$ in minisuperspace by:
\begin{align}
&u = {e^\xi }{e^\eta }\nonumber \\
&v = {e^\xi }{e^{-\eta }}
\end{align}
For the case of 10d nonsupersymmetric string theory this gives:
\begin{align}
&{e^\eta } = {a^{16}}{e^{3\phi /4}}\nonumber \\
&{e^\xi } = {a^9}{e^{5\phi /4}}
\end{align}
The Hamiltonian constraint in these coordinates is given by:
\begin{equation}H = {e^{ - 2\xi }}\left( { - p_\eta ^2 + p_\xi ^2} \right) + 1 = 0\end{equation}
The solutions to the WDW equation $H\psi=0$ in terms of the Rindler like $(\xi,\eta)$ minisuperspace variables are the Rindler wave functions which can be expressed in terms of the modified Bessel function of the second kind $K_\nu(z)$ through:
\begin{equation}{\psi _k}^{ \pm (3)}(\eta ,\xi ) = {K_{ik}}\left( {\sqrt \lambda  {e^\xi }} \right){e^{ \pm ik\eta }}\end{equation}
in terms of the $(a,\phi)$ this is:
\begin{equation}\psi _k^{ \pm (3)}(a,\phi ) = {\left( {{a^{15}}} \right)^{ \pm ik}}{e^{ \pm ik3\phi /4}}{K_{ik}}\left( {\sqrt \lambda  {a^9}{e^{5\phi /4}}} \right)\end{equation}
The general solution can be written as a superposition of the form:
\begin{equation}\psi (a,\phi ) = \sum\limits_k {\left( {A_k^{(3)}\psi _k^{ + (3)}(a,\phi ) + B_k^{(3)}\psi _k^{ - (3)}(a,\phi )} \right)} \end{equation}

%\section{11d gravity, two sided black holes and topology change}

\begin{figure}
\centering
  \includegraphics[width = .75 \linewidth]{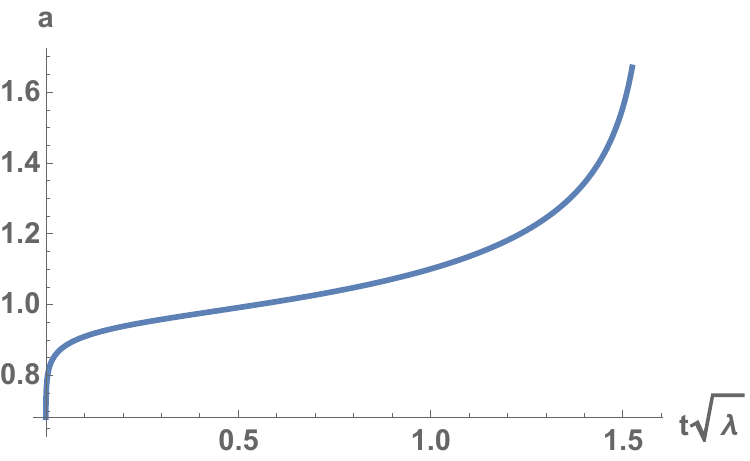}
  \caption{Scale factor $a$ for the 10d nonsupersymmetric string cosmological solution. The scale factor $a$ diverges at $t=0$ and at $t=\pi/(2\sqrt{\lambda})$.}
  \label{fig:Radion Potential}
\end{figure}
\begin{figure}
\centering
  \includegraphics[width = .75 \linewidth]{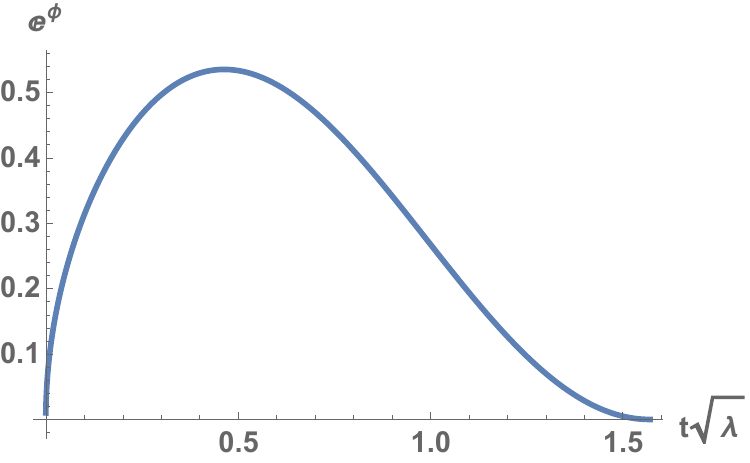}
  \caption{Exponential of the dilaton field for the 10d nonsupersymmetric string cosmological solution. The exponential of the dilaton goes to zero at $t=0$ and at $t=\pi/(2\sqrt{\lambda})$.}
  \label{fig:Radion Potential}
\end{figure}

\section{Inflation, dark matter and nonsupersymmetric string}

Most of the above discussion has been in 2d or 10d spacetimes or in 4d spacetime without matter. More realistic models are in 4d with different matter content of the standard model and its extensions. The theory of inflation introduces a scalar particle, the inflaton, with experimental signatures realted to the Cosmic Microwave Background (CMB)  data. This data can be used to constrain  the shape of the inflationary potential as well as the mass of the inflaton. Another particle present in realistic models is the Higgs boson whose mass was measured at the LHC. Still another set of 
particles are those that make up the dark matter. Nonsupersymmetric string theory compactifield on orbifolds provides models with Higgs like fields as well as candidates for the inflaton and dark matter. Also these theories can be coupled to gravity quantum mechanically and have massive string modes which can have implications in the early Universe, when the energy density was high enough to produce them. In this section we discuss these models from the point of view of canonical gravity in order to connect the previous discussion to more realistic models.  

\subsection*{Inflation}

The action for single field inflation is given by:
\begin{equation}S = \int {d^4{x}\sqrt { - g} } \left( {\frac{R}{{16\pi G}} - \frac{{{{\left( {\partial \phi } \right)}^2}}}{2} - V(\phi )} \right)\end{equation}
 $V(\phi)$ is the inflationary potential which leads to different predictions for the cosmological observations. Various inflationary models such as $R+R^2$ gravity, no-scale supergavity and Higgs inflation give rise to Starobinsky-like potentials \cite{Starobinsky:1980te}
\cite{Kinney}\cite{Cicoli:2023opf}
\cite{Ellis:2013xoa}
\cite{Brinkmann:2023eph}
\cite{Blumenhagen:2015qda}
\cite{Pallis:2016mvm}
\cite{Kaneta:2019yjn}. The prototype Starobinsky ptential is given by:
\begin{equation}V(\phi ) = \frac{3}{4}{\mu ^2}M_P^2{\left[ {1 - \exp \left( {\sqrt {\frac{2}{3}} \frac{\phi }{{{M_P}}}} \right)} \right]^2}\end{equation}
with the reduced Planck mass 
 ${M_P} = \frac{1}{{\sqrt {8\pi G} }} = 2.4 \times {10^{18}}GeV$ and the inflaton mass
 $\mu  = 3 \times {10^{13}}GeV $ taken to agree with astrophysics measurements of density perturbations. We plot the Starobinsky potential in figure 11 and 12 where we show the potential using $GeV$ and reduced Planck mass units. 

 Introducing a boundary term involving the extrinsic curvature in the action to cancel second derivative terms and using the ansatz:
 \[d{s^2} =  - {N^2}(t)d{t^2} + {a^2}(t)\sum\limits_{i = 1}^3 {d{x_i}^2} \]
 \begin{equation}\phi = \phi(t)\end{equation}
 the Lagrangian becomes:
 \begin{equation}L =  - \frac{{3a{{\dot a}^2}}}{N} + \frac{{{a^3}{{\dot \phi }^2}}}{{2N}} - {a^3}N V(\phi )\end{equation}
 We can solve the Euler-Lagrange equation from this Lagrangian numerically. 
 The inflationary behavior of the potential originates from the slow roll down the left part of the potential until it hits the exponential wall on the right part of the potential. Finally the oscillatory behavior of the $\phi$ field originates from the field oscillating near the bottom of the potential. A solution is plotted in figure 13 that illustrates the oscillatory behavior of the inflaton near the bottom of the potential.

\begin{figure}
\centering
  \includegraphics[width = .75 \linewidth]{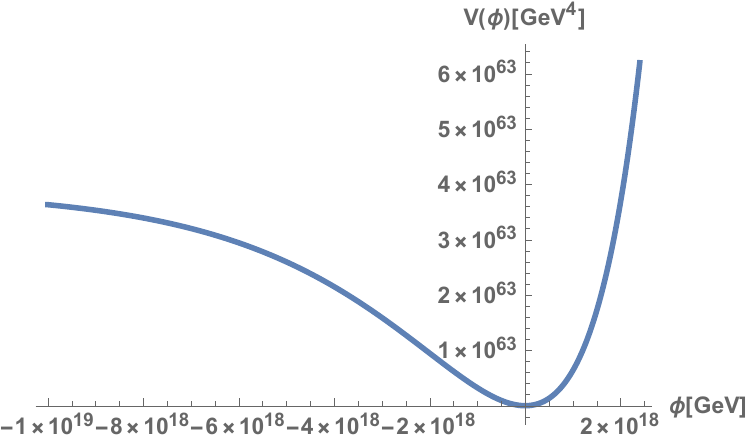}
  \caption{Starobinsky potential in units of $GeV^4$ for slow roll inflation. The slow roll region is to the left, while the potential well is at the origin, and exponential wall is to the right.}
  \label{fig:Radion Potential}
\end{figure}

\begin{figure}
\centering
  \includegraphics[width = .75 \linewidth]{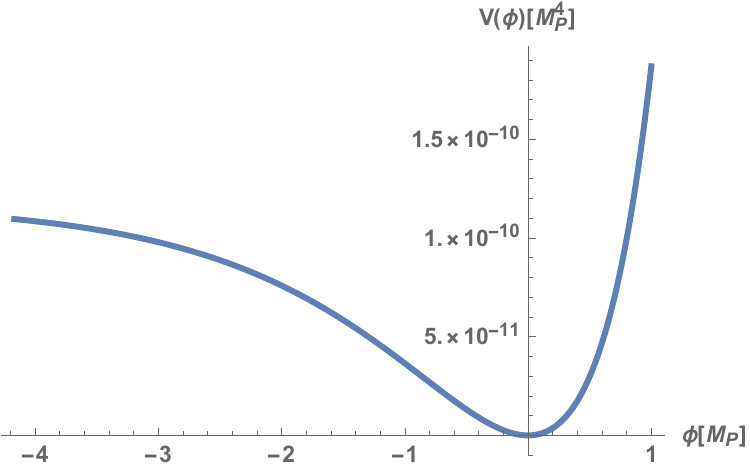}
  \caption{Starobinsky potential in units of $M_P^4$ for slow roll inflation where $M_P$ is the reduced Planck mass. The slow roll region is to the left, while the potential well at the origin, and exponential wall is to the right.}
  \label{fig:Radion Potential}
\end{figure}

\begin{figure}
\centering
  \includegraphics[width = .75 \linewidth]{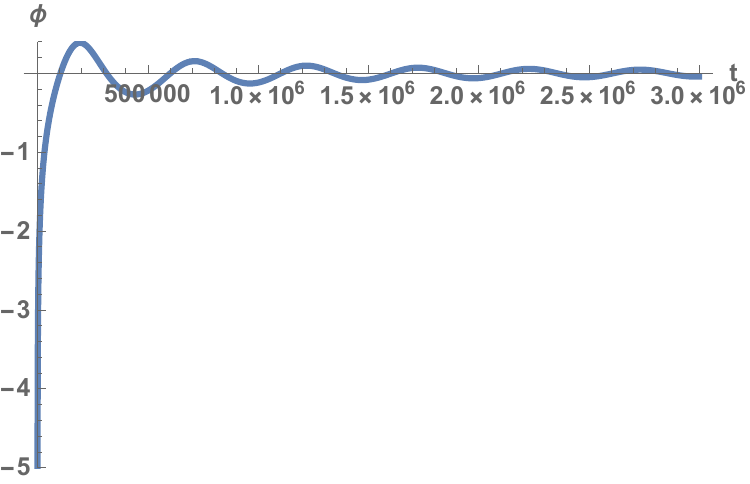}
  \caption{Inflaton  field  as a function of time for the Starobinsky potential. The initial conditions for the inflaton field and its derivative were chosen to illustrate the oscillation of the inflaton field at the bottom of the potential.}
  \label{fig:Radion Potential}
\end{figure}

To pursue the canonical gravity approach one forms the Hamiltonian constraint by varying the Lagrangian with respect to the Lapse function $N$. For the single field inflationary model we obtain:
\begin{equation}H =  - \frac{{3a{{\dot a}^2}}}{{{N^2}}} + \frac{{{a^3}{{\dot \phi }^2}}}{{2{N^2}}} + {a^3}V(\phi ) = 0\end{equation}
The canonical momentum associated with $a$ and $\phi$ from the Lagrangian are given by:
\begin{align}
&{p_a} =  - \frac{{6a\dot a}}{N}\nonumber \\
&{p_\phi } = \frac{{{a^3}\dot \phi }}{N}
\end{align}
Now expressing the Hamiltonian constraint in terms of canonical momentum we have:
\begin{equation}H =  - \frac{{p_a^2}}{{12a}} + \frac{{p_\phi ^2}}{2} + {a^3}V(\phi ) - ak\end{equation}
where we have added a term proportional to the spatial curvature and $k=1,0,-1$ for positive, flat and negative spatial curvature. 
The WDW for the equation for the wave function is given by:
\begin{equation}H \psi = \left( - \frac{{p_a^2}}{{12a}} + \frac{{p_\phi ^2}}{2} + {a^3}V(\phi ) - ak\right)\psi=0\end{equation}
To represent this equation as a second order differential equation one makes the substitution:
\begin{align}
&{p_a} =  - i\frac{\partial }{{\partial a}}\nonumber \\
&{p_\phi} =  - i\frac{\partial }{{\partial \phi}}
\end{align}
Multiplying the Hamiltonian constrain by $a^{-3}$ we rewrite the Hamiltonian constraint as:
\begin{equation}a^{-3}H\psi  = \left( { - \frac{{p_a^2}}{{12{a^4}}} + \frac{{p_\phi ^2}}{{2{a^6}}} - \frac{k}{{{a^2}}} + V\left( \phi  \right)} \right)\psi  = 0\end{equation}
The Starobinsky potential leads to slow roll inflation and can be compared directly to experimental data. The form of the potential is the same as  the Morse potential used to describe quantum molecules \cite{Morse:1929zz}.
The Morse potential is soluble quantum mechanically using supersymmetric quantum mechanics \cite{Apanavicius:2021yin}. The relation of the Starobinsky potential to the Morse potential suggests that Supersymmetric Quantum Mechanics (SUSYQM) methods \cite{Gangopadhyaya:2017wpf} my be useful to solve for the wave function the Starobinsky potential. For quantum cosmology with the Starobinsky potential the SUSYQM approach yields the Dirac square root of the WDW equation given by \cite{vanHolten:2017dok}
\cite{Mallett:1995ut}
\cite{Kim:1996yv}
\cite{DEath:1993xgi}
\cite{Yamazaki:2001ef}
\cite{Bogers:2015fwa}
\cite{Kan:2021fmw}
\cite{Kan:2022ism}
\cite{Balcerzak:2023elf}:
\begin{equation}\left( {{\gamma ^m}\left( {{E^{ - 1}}} \right)_m^a{p_a} + {\gamma ^m}\left( {{E^{ - 1}}} \right)_m^\phi {p_\phi } + \alpha W} \right)\psi  = 0\end{equation}
which is a first order equation similar to the supercharge acting of the wave function in SUSYQM. Here the the inverse minisuperspace vierbein associated with the minisuperspace metric is
\begin{equation}\left( {{E^{ - 1}}} \right)_m^a = \left( {\begin{array}{*{20}{c}}
{\frac{1}{{\sqrt {12} }}\frac{1}{{{a^2}}}}&0\\
0&{\frac{1}{{\sqrt 2 }}\frac{1}{{{a^3}}}}
\end{array}} \right)\end{equation}
and the superpotential $W$ of SUSQM
\begin{equation}W = \sqrt {\frac{3}{4}} \mu {M_P}\left[ {1 - \exp \left( {\sqrt {\frac{2}{3}} \frac{\phi }{{{M_P}}}} \right)} \right]\end{equation}
where  $\alpha, \gamma^i$ are Dirac matrices obeying: 
\[{\alpha ^2} = 1\]
\begin{equation}\alpha {\gamma ^m} + {\gamma ^m}\alpha  = 0\end{equation}
Another interesting application of the first order WDW-Dirac equation is to dark energy where in that case $W=\sqrt{\lambda}$ where $\lambda$ is the four dimensional cosmological constant \cite{vanHolten:2017dok}. This acts like a mass term in the Dirac equation and one can formulate an approach to small values of dark energy in a similar way to how one approaches small values of mass. If minisuperspace has a notion of chiral symmetry breaking for example one can obtain a suppression of the cosmological constant.

Unlike the cases considered above there does not yet exist an exact minisuperspace solution to the WDW equation for the Starobinsky potential. For the left part of the potential where $V(\phi)$ is slowly varying with respect to $\phi$ one can derive a WKB like wave function for the Starobinsky potential and from those obtain the Hartle Hawking and Vilenkin tunneling form of the wave function. These are given by \cite{Wiltshire:1995vk}:
\begin{align}
&{\psi _{WKB}} = \frac{{B(\phi )}}{{a{{({a^2}V(\phi ) - k)}^{1/4}}}}\exp \left[ {\frac{{ \pm i}}{{3V(\phi )}}{{({a^2}V(\phi ) - k)}^{3/2}}} \right]\nonumber \\
&{\psi _{HH}} = \exp \left( {\frac{1}{{3V(\phi )}}} \right)\exp \left[ {\frac{{ - 1}}{{3V(\phi )}}{{(1 - {a^2}V(\phi ))}^{3/2}}} \right]\nonumber \\
&{\psi _{T}} = \exp \left( {\frac{{ - 1}}{{3V(\phi )}}} \right)\exp \left[ {\frac{{ - i}}{{3V(\phi )}}{{({a^2}V(\phi ) - 1)}^{3/2}} + \frac{{i\pi }}{4}} \right]
\end{align}
For the region at the bottom of the Starobinsky potential the effective cosmological constant is zero and the scale factor is no longer rapidly varying. In this region one can treat $a$ adiabatically in the wave function and use the SUSYQM method to obtain the wave function using the relation to the Morse potential. Thus in the region near the bottom of the potential we can approximate the wave function as:
\begin{equation}{\psi _{Morse}} = {e^{{a^3}\sqrt {\frac{3}{2}} \mu {M_P}\phi }}\exp \left[ { - {a^3}\frac{3}{2}\mu M_P^2{e^{\sqrt {\frac{2}{3}} \frac{\phi }{{{M_P}}}}}} \right]\end{equation}
To connect the Starobinsky potential with string theory one needs to identify a candidate for the inflaton. One possibility is to take the dilaton as the inflaton with its effective action generated for loop effects and it's interactions  with other fields. The effective action of the dilaton interacting with a fermion to three loop order was carried out in \cite{Cabo:2008sj}
\cite{Cabo:2010py}
\cite{Cabo:2011zz}.
Indeed the effective action of the dilaton interacting with a fermion field yields a potential of a similar shape to the Starobinsky potential and is of the form:
\begin{equation}V(\phi ) = \sum\limits_{m,n} {{e^{m\phi }}{\phi ^n}{c_{mn}}} \end{equation}
where $m=\{4,6,8\}$ and $n=\{0,1,2,3,4\} $ The mass of the dilaton calculated from the second derivative of the effective potential is $6.93 \times 10^{19} GeV$ for a GUT scale mass fermion or $8.77 \times 10^{16} GeV$ for weak scale mass fermion. In either case it is too massive to be the inflaton associated with the Starobinsky potential. However one can include the interaction with other fields such as gauge bosons or scalar fields which could potentially lower the dilaton mass to align with astrophysical constraints on the mass of the inflaton.

\subsection*{Dark matter}

Another area of cosmology of great cosmological importance is dark matter which dominates over ordinary matter at cosmological scales and plays a fundamental role for structure formation in the Universe. There are several potential candidates for dark matter.
For supersymmetric models one has the Lightest Supersymmetric Particle (LSP) which can serve as a dark matter candidate of the Weakly Interacting Massive particle (WIMP) type. For nonsupersymmetric string theory one has $Z_2 \times Z_4$ orbifold compactifications of the 10d nonsupersymmetric string \cite{Perez-Martinez:2021zjj}  with a singlet scalar field in the hidden sector that can serve as a dark matter candidate \cite{Cervantes:2023wti}. This dark scalar field $s$ interacts with the Higgs field as a portal field. The Higgs potential in the standard model is given by \cite{Melo:2017agn}:
\begin{equation}{V_{Higgs}}(h) = \lambda {v^2}{h^2} + \lambda v{h^3} + \frac{\lambda }{4}{h^4}\end{equation}
where $h$ is the Higgs field $v = 246GeV$ and 
$ \lambda  = .13$. The  portal interaction  of the Higgs field with the with the hidden scalar $s$ is:
\begin{equation}{V_{dark}}(s,h) = \frac{1}{2}{y^2}{v^2}{s^2} + {y^2}v{s^2}h + \frac{1}{2}{y^2}{s^2}{h^2} + \frac{{\lambda '}}{4}{s^4}\end{equation}
where $y$ is the portal coupling and $\lambda'$ is the coupling constant for the self interaction of the hidden $s$ field. 

Another potential dark matter candidate for nonsupersymmetric string theory is a dark glueball coming from the gauge group in the hidden sector \cite{Faraggi:2000pv}
\cite{Soni:2016gzf}
\cite{Carenza:2023shd}
\cite{Curtin:2022oec}
\cite{Carenza:2022pjd}
\cite{Forestell:2017wov}
\cite{Acharya:2017szw}. This type of dark matter candidate is self interacting and and spinless for the lightest dark glueball. The dark glueball can suffer from overabundance during inflation if coupled to the inflaton. It becomes  more viable as a dark matter candidate satisfying astrophysical constraints if the electroweak  Higgs potential can be stabilized.  In particular if the Higgs couples to an additional scalar field, as in the dark sector $s$ field as above, the Higgs can be stabilized and the dark glueball avoids overabundance in the Starabinsky model of inflation \cite{Li:2021fao}
\cite{Adshead:2019uwj}
\cite{Lebedev:2012zw}
\cite{Elias-Miro:2012eoi}
\cite{Ema:2016ehh}. The derive the wave function for this cosmological scenario we start with the Lagrangian for doublet Higgs field given by:
\begin{equation}L_{Higgs} = {\left| {D_\mu }\mathcal{H} \right|^2} + \lambda {v^2}{\left| \mathcal{H} \right|^2} - \lambda {\left| \mathcal{H} \right|^4}\end{equation}
The doublet Higgs field can be written as:
\begin{equation}\mathcal{H} = \frac{1}{{\sqrt 2 }}\left( \begin{array}{l}
0\\
v + h
\end{array} \right)\end{equation}
To define a portal interaction with the hidden glue sector define dark gluon field strength $F'$ and write:
\begin{equation}\frac{c}{{M'}^2}\mathcal{H}^{\dagger}\mathcal{H} tr(F'F')\end{equation}
expanding out to leading order in $h$ we have:
\begin{equation}\frac{cvh}{{M'}^2} tr(F'F')\end{equation}
For minisuperspace and $SU(2)$ hidden gauge field we can use the ansatz for the gauge field and field strength given by \cite{Maleknejad:2012fw}
\cite{Emoto:2002fb}:
\begin{align}
&{A'}_i^a = \varphi (t)\delta _i^a\nonumber \\
&{F'}_{0i}^a = \dot \varphi (t)\delta _i^a\nonumber \\
&{F'}_{ij}^a =  - g{\varphi ^2}(t)\varepsilon _{ij}^a
\end{align}
where $g$ is the hidden gauge coupling constant. The total Lagrangian for the $(a,\phi,h,s,\varphi) $ minisuperspace is:
\begin{equation}L_{tot} =  - \frac{{3a{{\dot a}^2}}}{N} + \frac{{{a^3}\left( {{{\dot \phi }^2} + {{\dot h}^2} + {{\dot s}^2}} \right)}}{{2N}} + \frac{{3a{{\dot \varphi }^2}\left( {1 + \frac{{cvh}}{{M'}}} \right)}}{{2N}} - \frac{{3{g^2}{\varphi ^4}\left( {1 + \frac{{cvh}}{{M'}}} \right)}}{{2a}} - {a^3}\left( {V(\phi ) + {V_{Higgs}}(h) + {V_{dark}}(s,h)} \right)\end{equation}
and the canonical momentum derived from the Lagrangian are:
\[{p_a} =  - \frac{{6a\dot a}}{N}\]
\[{p_\phi } = \frac{{{a^3}\dot \phi }}{N}\]
\[{p_h} = \frac{{{a^3}\dot h}}{N}\]
\[{p_s} = \frac{{{a^3}\dot s}}{N}\]
\begin{equation}{p_\varphi } =  - \frac{{3a\dot \varphi \left( {1 + \frac{{cvh}}{{M'}}} \right)}}{N}\end{equation}
The Hamiltonian constraint in terms of the canonical momentum is:
\begin{equation}H_{tot} =  - \frac{{p_a^2}}{{12a}} + \frac{{\left( {p_\phi ^2 + p_h^2 + p_s^2} \right)}}{{2{a^3}}} + \frac{{p_\varphi ^2}}{{6a\left( {1 + \frac{{cvh}}{{M'}}} \right)}} + \frac{{3{g^2}{\varphi ^4}\left( {1 + \frac{{cvh}}{{M'}}} \right)}}{{2a}} + {a^3}\left( {V(\phi ) + {V_{Higgs}}(h) + {V_{dark}}(s,h)} \right)\end{equation}
So the WDW equation is:
\begin{equation}H_{tot}\psi(a,\phi,h,s,\varphi)=0  \end{equation}
Going beyond minisuperspace so that $A'$ is a function of spatial coordinates yields a Hamiltonian constraint equation of the from:
\begin{equation}H_{tot}\psi (a,\phi ,h,s,A') = 0\end{equation}
Solving this Hamiltonian equation is difficult. For the Hamiltonian  for the dark sector only we have:
\begin{equation}H_{dark}\psi (s,A') = 0\end{equation} 
For hidden groups such as pure $SU(N)$ Yang-Mills solving this equation for the wave function will require extensive computing resources. Quantum simulations in lower dimensions, discrete gauge groups or effective matrix models can serve as benchmarks towards this goal \cite{Butt:2022xyn}
\cite{Miceli:2019snu}
\cite{Pisarski:2021aoz}
\cite{Desai:2021oiy}
\cite{Feng:2021mdi}.

\subsection*{String aspects}

Although most of the previous discussion is in the context of effective actions from nonsupersymmetric string theory, more direct effects from string theory may play a role, especially near singularities where energy and pressure densities grow to Planckian values \cite{Maldacena}
\cite{Arkani-Hamed:2015bza}. A canonical gravity approach to direct string effects is hampered by the lack of an analog of the Wheeler De Witt equation for string theory \cite{Banks:1986kd}. This an old problem dating back to the early days of string theory and is related to the lack of a manifestly background independent formulation of the theory as well as the problem of a canonical formulation with higher time derivatives from the massive fields and nonlocality of string interactions. Still one can investigate the effect of the lowest massive states of the nonsupersymmetric string theory. For 10d nonsupersymmetric string theory the lowest massive fields $B$ and $C$ are of the form \cite{Green:1987mn}:
\begin{equation}(36;128,16') + (36;16,128')\end{equation}
where $36$ are massive antisymmetric tensor indices, $128,128'$ are $SO(16),SO(16)'$ spinor indices and $16,16'$ are $SO(16),SO(16)'$ vector indices The lowest massive modes have $147,456$ degrees of freedom and  mass terms given by
\begin{equation}\frac{4}{2\alpha'}B_{\mu \nu }^{AI'}B_{\mu \nu }^{AI'} + \frac{4}{2\alpha'}C_{\mu \nu }^{IA'}C_{\mu \nu }^{IA'}\end{equation}
 It is interesting that these massive fields carry the quantum numbers both the visible and hidden sectors and can serve as portal fields connecting the two sectors and generating effective interaction terms at low energies. Unitarity of string theory requires a tower of massive states to be included and these can have different interactions with the curvature tensor than what would be expected in field theory \cite{Giannakis:1998wi}. Higher massive modes have an enormous number of degrees of freedom and behave as statistical systems with their decay to massless modes at the Hagedorn temperature. Thus the decay of these string modes can yield a high temperature yet homogeneous equation of state in the early Universe with  Hagedorn initial temperature \cite{Gross:2021gsj}
\cite{Wilkinson:1989tb}
\cite{Maggiore:1997vw}
\cite{Kawamoto:2013fza}. For these high tempratures black holes can be thermally produced and the final stage of evaporation involves Planckian physics can also involve string considerations \cite{Chen:2021dsw}.

String duality can also play a role and one can map initial states at large $a$ to final states with small $a$ \cite{Giveon:1994fu}
\cite{Rocek:1991ps}
\cite{Gasperini:2002bn}
\cite{Martinec:1994xj}
\cite{Lawrence:1995ct}
\cite{Sen:1991zi}
\cite{McGuigan:1990pi}. String duality transformations can take one to solutions of totally different string theories. This suggests that M-theoretic considerations are necessary for Planckian values of the scale factor near the singularity. Like string theory  M-theory  also lacks a background independent formulation and instead can be defined holographically using Matrix models \cite{Banks:1996vh}.  To explore the implications of M-theoretic aspects of cosmology Hamiltonian simulation for Matrix formulations of M-theory can be developed but are  computational intensive on classical and quantum computers \cite{Maldacena:2023acv}
\cite{Chandra:2022mae}
\cite{Rinaldi:2021jbg}. Another application of Holography in cosmology relates holography to canonical gravity \cite{Raju:2019qjq}
\cite{Chakraborty:2023los} and de Sitter space \cite{Chakraborty:2023yed}
\cite{Blacker:2023oan} which may have implications to dark energy and inflation which is approximately de Sitter during the inflationary phase. Finally another approach to canonical gravity which is inspired by string theory is to include ghost fields in the wave function \cite{Witten:2022xxp}
\cite{Ljatifi:2023dro}. The Hamiltonian constraint is then replaced by annihilation of the wave function by the BRST charge and BRST Laplacian \cite{VanHolten:2001nj}
\cite{vanHolten:1995ds}. One can expand the wave function in the ghost field and define an inner product by Grassmann integration which is equivalent to the Klein-Gordon inner product of the wave function of DeWitt. In the case when the metric depends on time and one spatial coordinate, midisuperspace, the nilpotency of BRST charge is related the vanishing of beta functions defined through midisuperspace coordinates \cite{McGuigan:1990nd}
\cite{Cooper:1991vg}.

\section{Conclusion}

In this paper we have investigated a quantum cosmology approach to black hole interiors using canonical gravity and nonsupersymmetric string theory. We find that the same techniques are useful for both string cosmology and black hole interiors and even the wave functions are identical for certain choices of coordinates in minisuperspace. We investigated models in 2d, 4d, 10d and more realistic models involving orbifold compactifications of nonsupersymmetric string theory. In particular we investigated one model which included cosmic inflation with a Starobinsky like potential, a Higgs sector, a hidden scalar field and a dark gluon field. We also discussed string aspects of these models, in particular those associated with massive string modes. It will be interesting to study these aspects of the model in more detail to see what modifications to the predictions of cosmic inflation and dark matter cosmology can be obtained, and if improvements of cosmological measurements can be made precise enough to yield information on which form of Starobinsky like potential is most favored. Finally measurements of primordial gravitational radiation could help determine the phase structure and number of degrees of freedom   \cite{Matzner}
\cite{Borah:2021ftr} which can also help pin down the type of model which best describes the early Universe. From both the theory and experimental perspective it is  an exciting time to study cosmology.

\end{document}